\documentclass[10pt,twocolumn]{IEEEtran}
\usepackage{graphicx,latexsym}
\newcommand{\figura}[3]
{
\begin{figure}
  \centering
 \includegraphics[width=8cm]{#1}
  \caption{#2}\label{#3}
\end{figure}
}

\title{Connection Between System Parameters and Localization Probability in
Network of Randomly Distributed Nodes}
\author{F.~Daneshgaran, \IEEEmembership{Member,~IEEE,}
M.~Laddomada, \IEEEmembership{Member,~IEEE,} and 
M.~Mondin,~\IEEEmembership{Member,~IEEE}
\thanks{\textbf{To appear on IEEE Transactions on Wireless Communications, November 2007}}
\thanks{This work has been supported by NSF grant entitled "Hourglass: an Infrastructure
for Sensor Networks".}
\thanks{Fred Daneshgaran is with the ECE Dept.,
Calif. State Univ., Los Angeles, USA.}
\thanks{Massimiliano Laddomada (\textrm{laddomada@polito.it}) and Marina Mondin
are with the Dipartimento di Elettronica, Politecnico di Torino,
Italy.}}
\begin{document}
\maketitle
\begin{abstract}
This article deals with localization probability in a network of
randomly distributed communication nodes contained in a bounded
domain. A fraction of the nodes denoted as L-nodes are assumed to
have localization information while the rest of the nodes denoted
as NL nodes do not. The basic model assumes each node has a
certain radio coverage within which it can make relative distance
measurements. We model both the case radio coverage is fixed and
the case radio coverage is determined by signal strength
measurements in a Log-Normal Shadowing environment. We apply the
probabilistic method to determine the probability of NL-node
localization as a function of the coverage area to domain area
ratio and the density of L-nodes. We establish analytical
expressions for this probability and the transition thresholds
with respect to key parameters whereby marked change in the
probability behavior is observed. The theoretical results
presented in the article are supported by simulations.
\end{abstract}
%
%
\section{Introduction}
This article deals with a network of randomly distributed nodes some
of which have a-priori knowledge about their position (these are
the so called L-nodes) while others are supposed to localize
themselves (these are the so called non-localized or NL-nodes).

This general framework can be recognized in many practical
scenarios and in different types of communication networks, such
as Distributed Sensor Networks (DSNs) and wireless networks.
Generally speaking, we may classify the localization algorithms in
at least two ways, centralized or distributed \cite{neal} and
range-free or based on ranging techniques \cite{chong}. The most
common techniques are based on measured range whereby the location
of nodes are estimated through some standard methods such as
triangulation. The ranging technique itself can be based on: 1)
Received Signal Strength (RSS) measurements, 2) Time Of Arrival
(TOA) measurements requiring clock synchronization among nodes, 3)
Time Difference Of Arrival (TDOA) techniques requiring relative
synchronicity of L-nodes, 4) Angle Of Arrival (AOA) measurements,
and 4) hybrid combination of the above.

In range-free localization, connectivity between nodes is a binary event, either two nodes are within communication
range of each other or they are not \cite{neal2}.
For simplicity, we may view this as a hard quantization of for instance
RSS. If RSS is above a certain detection threshold, the nodes can communicate, otherwise they cannot. Of course,
the nature of path loss and the terrain characteristics influences both the coverage radius and the deviation
of the coverage zone from the idealistic circular geometry. Various range free algorithms have been proposed
in the literature including the centroid algorithm \cite{inchong1}, the DV-HOP algorithm \cite{inchong8},
the Amorphous positioning algorithm \cite{inchong7},
APIT \cite{inchong4}, and ROCRSSI \cite{chong}.
The fundamental characteristic of these algorithms is their ability to provide coarse
(compared to ranging techniques)
localization information with minimal per node communication and computation requirements.

Regardless of whether ranging based or range-free techniques are
used for localization, several global issues of concern remain to
be addressed in their application. In particular, this article deals
with broader questions of whether a given NL-node can localize
itself in a random array of nodes, a percentage of which have
localization information. What are the requirements on the density
of nodes over a terrain and coverage area of individual nodes that
would allow node localization with very high probability?

These questions and others like them are at the heart of the
localization problem and by their very nature, require a
probabilistic setting \cite{prob}. To have a frame of reference,
we present a communication scenario which is needed before any
localization of NL-nodes can take place. In particular, we assume:
1) ranging techniques are employed and the coverage radius of each
node is dependent on the path loss characteristic. While the
coverage radius may vary from node to node, we assume that the
coverage region is circular. The fact that the coverage geometry
may not be perfectly circular has no significant bearing on the
calculated probabilities. This is because of our assumption of
uniform distribution of nodes on the terrain whereby what matters
in so far as probability of finding other nodes in a given area
around a reference node is concerned, is the area itself and not
its geometric shape; 2) the nodes have calibrated power levels
possibly achieved via an external beacon signal, 3) a CSMA/CA link
layer protocol is employed for communication among nodes, hence,
while collisions might occur, we can assume that the nodes within
communication range can communicate without difficulty, and 4)
periodic broadcasts by L-nodes are used to inform NL-nodes that
can hear them of the coordinate location of the L-nodes relative
to some absolute reference frame.

A variety of localization techniques for both indoor and outdoor
environments are currently available, offering various trade-offs
between accuracy, cost and complexity. In this article we assume
that each NL-node needs to communicate with at least three other
L-nodes in order to be able to localize itself. A review of
various localization techniques proposed in the literature may be
found in \cite{Bulusu}. In \cite{Shang}, the authors propose an
approach based on connectivity information, to derive the
locations of nodes in a network. In \cite{Chen}, the authors
present some work in the field of source localization in sensor
networks.

The rest of the article is organized as follows. In section-II we present the theoretical
analysis for the localization probability. Section-III
is devoted to presenting the simulation results in support of the theoretical analysis and section-IV
to conclusions.
\section{Analysis of Localization Probability Using the Probabilistic Method}
The basic parameter set defining the problem is as follows.
\begin{itemize}
\item Total number of nodes distributed uniformly over a circular
disk of radius $R$ (denoted as the domain) is $n$.
\item $k$ of the $n$ nodes are assumed to be L-nodes (implying
they have localization information relative to some coordinate
frame. How this localization is established is irrelevant to
problem formulation), the rest are denoted as non-localized nodes
(NL-nodes) and need to localize themselves.
\item The radio coverage radius is denoted as $d$ and is defined
by $d=\sqrt{A_{cov}/\pi}$ (we assume $d<<R$). In particular, the
shape of the coverage area $A_{cov}$ is irrelevant in so far as
the calculation of the probabilities is concerned due to the
assumption of the uniform distribution of the nodes in the
terrain.
\item The localization problem is two dimensional and three
distance measurements relative to nodes with known positions is
sufficient to solve for the $(X,Y)$ coordinates of the NL-node
unambiguously.
\item We shall neglect the boundary problem in the sense that the
nodes near the boundary of the domain can only see other network
elements within the domain. Hence in effect, their radio coverage
area within which they may identify other nodes is reduced. This
assumption is validated to be reasonably good, even when the total
number of nodes is not very large.
\end{itemize}
Various localization techniques have been reported in the
literature. The basic strategy assumed in this article is as follows
\cite{crespi1}.
\begin{enumerate}
\item The graph corresponding to the set of nodes is created with
bidirectional arcs based on range. In particular, any pair of
nodes whose true distances are within a certain limit are
connected by a bi-directional edge and can communicate with each
other.
\item A given NL-node that has not localized itself, stores two
sets of information generated locally: i) absolute coordinates of
the L-nodes within its radio coverage (if any), ii) distance
estimates to all other nodes within its radio coverage based on
mean power measurements. We shall assume the measured distance
information to be accurate.
\end{enumerate}
The basic question that we wish to answer in this article is as
follows; given that we are at a NL-node that is interior to our
domain, what is the probability that the node can localize itself?
To localize itself a NL-node needs {\em at least} three L-nodes
within its radio coverage. Given the existence of degenerate
cases, the formulas presented in the Theorems below really
represent lower bounds to the node localization failure
probability. However, since the degenerate cases are very rare,
the calculated probabilities are very close to the true values
(i.e., the bound is tight), and we have verified this experimentally.\\

\noindent {\bf Theorem~2.1:} {\em Under the assumption of a
uniform distribution of $n$ nodes, $k<n$ of which are L-nodes
while the rest are NL-nodes, over a circular domain of radius $R$,
and per node radio coverage radius of $d<R$, the NL-node
localization failure probability is tightly lower bounded by:
\begin{equation}\small
\begin{array}{lll}
P_F &\geq& \left\{ \frac{(1-a)^2}{2}\frac{\partial^2}{\partial
a^2} +
(1-a)\frac{\partial}{\partial a} + 1 \right\} \circ \left\{ [b^2  a+(1-b^2 )]^{n-1}\right\} \\
   &=&  [1-(1-a)\cdot b^2 ]^{n-3}\cdot \left[1+b^2
   (1-a)(n-3)+\right.\\
   &&\left.+b^4 (1-a)^2 \frac{(n-1)(n-2)}{2} \right]\\
\end{array}
\end{equation}
where $a=(1-\frac{k}{n})$ (i.e., the fraction of the NL-nodes),
$b=\left( \frac{d}{R}\right)$ and the operator notation of vector
calculus is used to simplify the expression.
}\\

\noindent {\em Proof.} The assumption of uniform distribution of
nodes over our domain implies that the probability of having a
node within a circular disk (i.e., the radio coverage zone of a
given node) of radius $d$ is given by $\left(
\frac{A_{cov}}{A_{domain}}\right) =\left( \frac{d}{R}\right) ^2$.
In $p$ independent trials, the probability of having $p$ nodes
within the coverage zone is simply $\left( \frac{d}{R}\right)
^{2p}$. Probability of a NL-node localization failure is bounded
by:
\begin{equation}
\begin{array}{llll}
P_F &\geq& P\{ p\leq 2 \; or \; (p\geq 3 \; \mbox{and at most two
nodes}\\
&& \mbox{are L-nodes while the rest are NL-nodes})\}
\end{array}
\end{equation}
The probability that $p=0$, is the probability that all other
$(n-1)$ nodes are outside the coverage zone of the NL-node and is
given by $[1-b^2 ]^{n-1}$. Similarly, the probability that $p\geq
1$ nodes are within the coverage radius is:
\[
\left( \begin{array}{c} n-1 \\ p \end{array}\right) b^{2p}[1-b^2
]^{n-1-p}
\]
If in addition, we require that of the $p$ nodes {\em at most} two
be L-nodes while the other $(p-2)$ nodes are NL-nodes, then the
probability of the event of interest is:
\[
\begin{array}{ll}
\left( \begin{array}{c} n-1 \\ p \end{array}\right) b^{2p}[1-b^2
]^{n-1-p}&\left[  a^p +pa^{p-1}(1-a)+\right.\\
&\left. +\frac{p(p-1)}{2}a^{p-2}(1-a)^2 \right]
\end{array}
\]
Putting the pieces together, after some algebra the NL-node
localization failure probability can be written as (assuming
$n\geq 4$):
\begin{equation}
\begin{array}{lll}
P_F &\geq &\sum_{p=0}^{(n-1)}\left( \begin{array}{c} n-1 \\
p
\end{array}\right) b^{2p}[1-b^2 ]^{n-1-p} \left[a^p +\right.\\
&& \left. +pa^{p-1}(1-a)+ \frac{p(p-1)}{2}a^{p-2}(1-a)^2 \right]
\end{array}
\end{equation}
This expression can further be written in the form:
\begin{eqnarray}
P_F&\geq&\frac{(1-a)^2}{2}\frac{\partial^2}{\partial a^2}\left\{
[b^2 \cdot a+(1-b^2 )]^{n-1}\right\} +\nonumber
\\
&&+(1-a)\frac{\partial}{\partial a}\left\{ [b^2\cdot a+(1-b^2
)]^{n-1}\right\} + \nonumber\\
&& \left[b^2 \cdot a+(1-b^2 )\right]^{n-1}
\end{eqnarray}
$\Box$
\\
When $(1-a)\cdot b^2 <<2/n$ we have the approximation:
\begin{equation}
P_F \simeq 1-\left[(n-3)\cdot (1-a)\cdot b^2 \right]^2
\end{equation}
Fig.~\ref{fig1} depicts the localization probability $(1-P_F )$ as
a function of the fraction of the NL-nodes, $a$, for exponentially
increasing values of $b$ defined via the expression
$b(j)=2^{j/22}-1, \; j=3,4,...,20$ in a network containing $n=300$
nodes. It is evident from the figure that the localization probability
increases as the fraction of the NL-nodes decreases for a fixed $b$, and for a given
fraction of NL-nodes, the localization probability increases rapidly as a function
of increasing $b$.
\figura{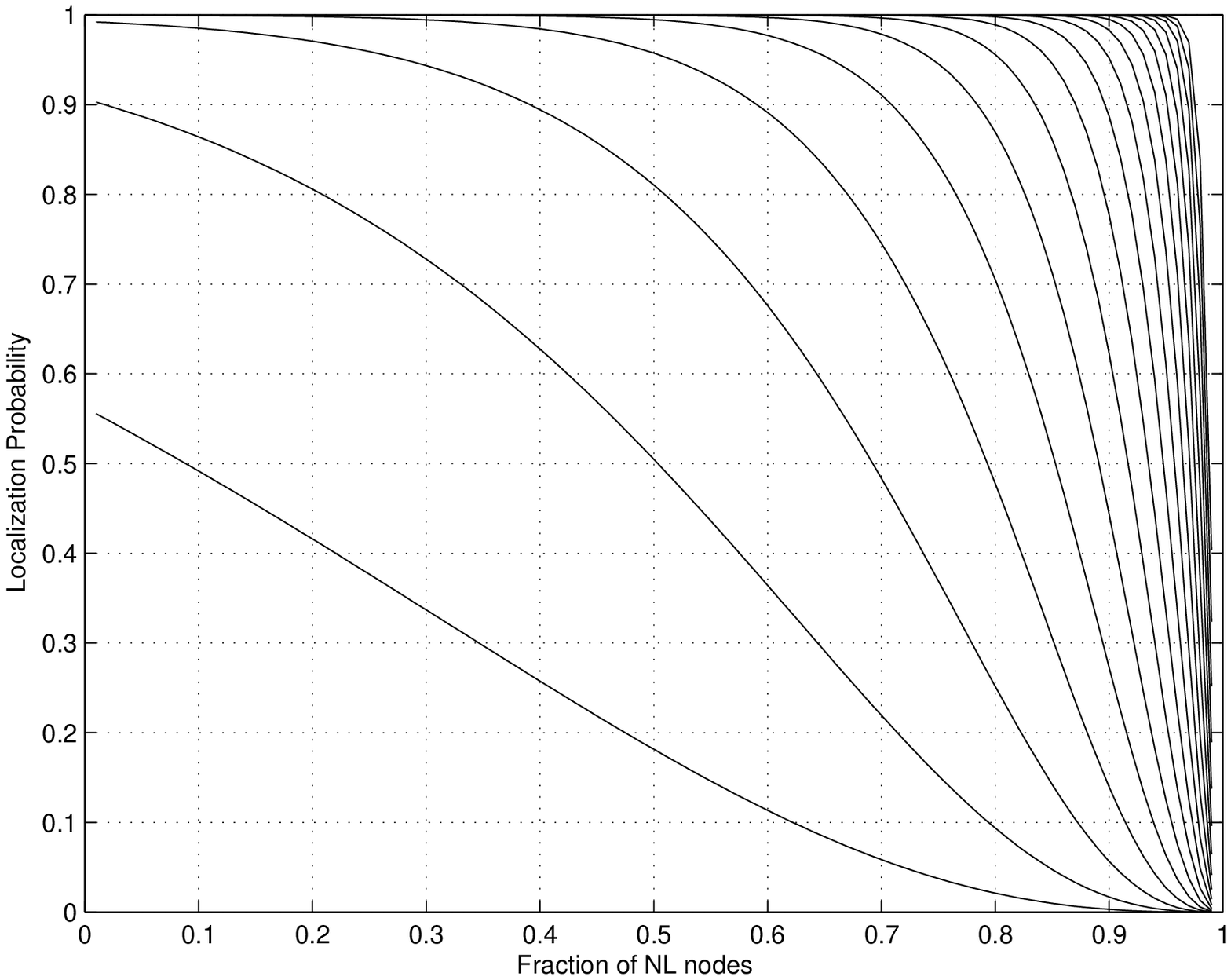}{Localization probability
as a function of $a$ for various values of $b$ for a network
containing $n=300$ nodes. The leftmost curve is associated with
the smallest $b$ ($b=0.099$), the rightmost curve is associated
with the largest $b$ ($b=0.878$), with $b$ varying as
$b(j)=2^{j/22}-1, \; j=3,4,...,20$. Increasing $b$ monotonically
increases the localization probability for a given~$a$.}{fig1}
\subsection{Extension to Log-Normal Shadowing}
Our analysis above on the node localization failure probability
may really be taken to represent a probability conditioned on two
things: 1) the total number of nodes $n$ is known, and 2) the
radio coverage area of the NL-node is known. In this subsection,
we wish to demonstrate that it is relatively straightforward to
remove these conditions. In particular, we shall take a particular
propagation model and remove the conditioning on the knowledge of
the coverage area. To this end, consider the problem of localizing
a specific node by signal power measurements. Let us consider the
received power $P(d)$ at a distance $d$ from a specific point
$
P(d)=P_o-10n_p \log_{10}(d)+X_s
$
whereby $P_o$ is the signal power at a reference distance $d_o$ normalized to one for simplicity,
$n_p$ is the path loss exponent, and $X_s$ is a
Gaussian-distributed random variable taking into account the
shadowing effect, i.e, $X_s\sim N(\mu_s,\sigma_s^2)$ with $\mu_s=0$.
An estimator of $d$ is given by $\widehat{d}=10^{\frac{P_o-P}{10n_p}}$, where
$P$ is the measured power.
With simple mathematical calculation, it is possible to obtain:
$
\widehat{d}=10^{\frac{P_o-P}{10n_p}}=10^{\log_{10}(d)-\frac{X_s}{10n_p}}=d\cdot
10^{-\frac{X_s}{10n_p}}
$
and then $\widehat{d}/d=10^{-\frac{X_s}{10n_p}}$.
The random variable $X_1=X_s/n_p$ is a gaussian random variable
$N(0,\sigma_1^2)$ with zero mean and variance
$\sigma_1^2=\sigma_s^2/ n_{p}^2$. We take $d$ to be the true
distance that would be measured if $X_s =0$. Hence, $X_s$ behaves
as a perturbation affecting the measured distance $\widehat{d}$.
Let the receiver have a detection threshold of $\gamma$. Hence, if
the received power is below $\gamma$, the receiver does not detect
the presence of a signal. This puts an upper limit on the
estimated distance as
$\widehat{d}_{max}=10^{\frac{P_o-\gamma}{10n_p}}$, leading to an
upper limit of $\frac{\widehat{d}}{d}\leq
\frac{\widehat{d}_{max}}{d}$. We note that if the received power
is below $\gamma$, the transmit and receive nodes do not see each
other. Hence the probability of finding a node in the coverage
zone of the transmit node is zero. Since our probability values
are proportional to the size of the coverage area, we interpret
this as the event that $\hat{d}=0$.

With this setup, the ratio between the estimated distance (when
measurable) and actual one $d$, i.e., $Y=\widehat{d}/d$ is a
random variable with log-normal probability density function
$f_Y(y)$ (initially we assume $\gamma=-\infty$ dB). In fact,
considering the following transformation applied to the random
variable $X_1\sim N(0,\sigma_1^2)$;
$
Y=g(X_1)=10^{-\frac{X_1}{10}}
$
it is simple to see that $\frac{\widehat{d}}{d}$ is distributed as
follows:
\begin{equation}\label{lognormal_distribution}
f_Y(y)=\frac{f_{X_1}(x_1)}{|g^{'}(x_1)|}_{|_{x_1=g^{-1}(y)}}=\frac{\alpha}{\sqrt{2\pi
\sigma_1^2}y}e^{\frac{-(10\cdot \log_{10} (y))^2}{2\sigma_1^2}}
\end{equation}
whereby $\alpha=\frac{10}{\ln (10)}$, $f_{X_1}(x_1)\sim
N(0,\sigma_1^2)$, and $|g^{'}(x_1)|=\frac{\ln
(10)}{10}10^{-\frac{x_1}{10}}$.

An important parameter in the analysis that follows is
the coverage to domain radius ratio. Consider the random
variable:
$
\frac{\widehat{d}}{R}=\widehat{b}=\left(\frac{\widehat{d}}{d}\right)\cdot
\left(\frac{d}{R}\right)
$
whereby $b_o=\frac{d}{R}$ is a fixed constant representing the
true coverage to domain radius ratio. With this setup, we have
$\widehat{d}/R=\widehat{b}=b_o\cdot Y$.
Define:
\[
f_1 (x)=\frac{\alpha}{\sqrt{2\pi
\sigma_1^2}x}e^{\frac{-\left(10\cdot \log_{10} (x)-10\cdot
\log_{10} (b_o)\right)^2}{2\sigma_1^2}}
\]
The probability density function (pdf)
$f_{\widehat{b}}(\widehat{b})$ of $\widehat{b}$ is:
\begin{equation}\label{lognormal_distribution_2}
\left\{ \begin{array}{ll} \frac{\alpha}{\sqrt{2\pi
\sigma_1^2}\widehat{b}}e^{\frac{-\left(10\cdot \log_{10}
(\widehat{b})-10\cdot \log_{10} (b_o)\right)^2}{2\sigma_1^2}}+&\\+
\left( \int_{\frac{\widehat{d}_{max}}{R}}^{\infty}f_{1} (x)dx
\right)
\delta(\widehat{b}), & 0\leq \widehat{b} \leq \widehat{b}_{max} \\
0 & \widehat{b} > \widehat{b}_{max} \end{array} \right .
\end{equation}
Note that $\widehat{b}$ is a mixed random variable. The
probability $P(\widehat{b}=0)\neq 0$. This follows from the
argument above that the event that the received signal is below
the detection threshold is equivalent to the event that
$\hat{d}=0$ implying $\hat{b}=0$. Again, considering the detection
threshold, we have the upper limit $\widehat{b}\leq
\widehat{d}_{max}/R$. To proceed further, we make the basic
assumption that $\frac{\widehat{d}_{max}}{R}< 1$ (this is almost
always true) and that the coverage region is entirely contained in
our domain.  Note that the detection threshold $\gamma$ and path
loss exponent have a major impact on $\widehat{d}_{max}$, but, we
always assume $R$ is large enough that the above condition
holds.\\

\noindent {\bf Theorem~2.2:} {\em Under the assumption of a
uniform distribution of $n$ nodes, $k<n$ of which are L-nodes
while the rest are NL-nodes, over a circular domain of radius $R$,
and per node radio coverage governed by log-normal shadowing with
$\frac{\widehat{d}_{max}}{R}<1$, the NL-node localization failure
probability $P_F$ is tightly lower bounded by:
\begin{equation}\small\label{closed_form_PF_1_2}
\sum_{l=0}^{n-3}\left(%
\begin{array}{c}
  n-3 \\
  l \\
\end{array}%
\right)\left(-k_1\right)^l\left\{E\left[\widehat{b}^{2l}\right]+k_2E\left[\widehat{b}^{2l+2}\right]+k_3E\left[\widehat{b}^{2l+4}\right]\right\}
\end{equation}
where $a=(1-\frac{k}{n})$ (i.e., the fraction of the NL-nodes),
$k_1=1-a$, $k_2=(1-a)\cdot(n-3)$, and $k_3=(1-a)^2\cdot
\frac{(n^2-3n+2)}{2}$.
}\\

\noindent {\em Proof.}
From Theorem~2.1 the NL-node localization
failure probability conditioned on $\widehat{b}$ can be written as
follows:
\[
P_{F|\widehat{b}}\geq\left[1-k_1\cdot
\widehat{b}^2\right]^{n-3}\cdot\left[1+k_2\widehat{b}^2+k_3\widehat{b}^4\right]
\]
Using the fact that:
\[
\left[1-k_1\cdot \widehat{b}^2\right]^{n-3}=\sum_{l=0}^{n-3}\left(%
\begin{array}{c}
  n-3 \\
  l \\
\end{array}%
\right)\left(-k_1\widehat{b}^2\right)^l
\]
it is possible to obtain the following relation:
\begin{equation}
P_{F|\widehat{b}}\geq\sum_{l=0}^{n-3}\left(%
\begin{array}{c}
  n-3 \\
  l \\
\end{array}%
\right)\left(-k_1\right)^l
\cdot\left[\widehat{b}^{2l}+k_2\widehat{b}^{2l+2}+k_3\widehat{b}^{2l+4}\right]
\label{pf_bhat}
\end{equation}
With this setup, the NL-node localization probability failure
$P_F$ can be lower bounded as follows:
\[
P_F=\int_0^{+\infty}P_{F|\widehat{b}}\cdot
f_{\widehat{b}}(\widehat{b})d\widehat{b}.
\]
By substituting (\ref{pf_bhat}) in the previous equation, it is
possible to obtain:
\begin{equation}
\begin{array}{lll}
P_F&\geq&\sum_{l=0}^{n-3}\left(%
\begin{array}{c}
  n-3 \\
  l \\
\end{array}%
\right)\left(-k_1\right)^l \left[
\int_0^{+\infty}\widehat{b}^{2l}f_{\widehat{b}}(\widehat{b})d\widehat{b}+\right.\\
&&\left. +k_2
\int_0^{+\infty}\widehat{b}^{2l+2}f_{\widehat{b}}(\widehat{b})d\widehat{b}+
k_3\int_0^{+\infty}\widehat{b}^{2l+4}f_{\widehat{b}}(\widehat{b})d\widehat{b}
\right]
\end{array}
\end{equation}
Note that the contribution of the delta function of $f_{\widehat{b}}(\widehat{b})$ in above integrals
is zero since we are looking at non-central moments.  This way a
closed form bound for $P_F$ can be written as follows:
\begin{equation}\label{closed_form_PF_1}
\begin{array}{lll}
P_F&\geq&\sum_{l=0}^{n-3}\left(%
\begin{array}{c}
  n-3 \\
  l \\
\end{array}%
\right)\left(-k_1\right)^l\left\{E\left[\widehat{b}^{2l}\right]+\right.\\
&&\left.+k_2E\left[\widehat{b}^{2l+2}\right]+k_3E\left[\widehat{b}^{2l+4}\right]\right\}
\end{array}
\end{equation}
$\Box$\\
Under certain conditions the expectations above are approximated by the
moments of a log-normal random variable $\widehat{b}$.
The k-th order moment of a log-normal random variable is given by,
\begin{equation}\label{moments}
E\left[y^{k}\right]=e^{\frac{k}{\alpha}\mu+\frac{1}{2}\left(\frac{k}{\alpha}\right)^2\sigma_1^2}
\end{equation}
whereby $\mu$ is the mean of the random variable $10 \log_{10}y$.
In particular, provided that the number of nodes $n$ is small
(less than 10) and the variance of shadow variable $X_s$ is
sufficiently small and recalling that $\alpha=\frac{10}{\ln 10}$
and $\mu=10\cdot \log_{10} (b_o)$, we have the approximation:
\begin{equation}\label{closed_form_PF_2}
\begin{array}{ll}
P_F\simeq & \sum_{l=0}^{n-3}\left(%
\begin{array}{c}
  n-3 \\
  l \\
\end{array}
\right)\left(-k_1\right)^l\left[
e^{\frac{2l}{\alpha}\mu+\frac{1}{2}\left(\frac{2l}{\alpha}\right)^2\sigma_1^2}+\right.\\
&\left. +k_2
e^{\frac{2l+2}{\alpha}\mu+\frac{1}{2}\left(\frac{2l+2}{\alpha}\right)^2\sigma_1^2}
+k_3
e^{\frac{2l+4}{\alpha}\mu+\frac{1}{2}\left(\frac{2l+4}{\alpha}\right)^2\sigma_1^2}
\right]
\end{array}
\end{equation}
The above expression can be written in the alternate form:
\begin{equation}\label{closed_form_PF_3}
\begin{array}{lll}
P_F&\simeq& \sum_{l=0}^{n-3}\left(%
\begin{array}{c}
  n-3 \\
  l \\
\end{array}
\right)\left(-k_1\right)^l\left[ b_{o}^{2l} e^{\frac{\sigma_1^2
(2l)^2}{37.72}}+\right.\\
&&\left. +k_2 b_o^{(2l+2)}e^{\frac{\sigma_1^2 (2l+2)^2}{37.72}}
+k_3 b_o^{(2l+4)}e^{\frac{\sigma_1^2 (2l+4)^2}{37.72}} \right]
\end{array}
\end{equation}
\figura{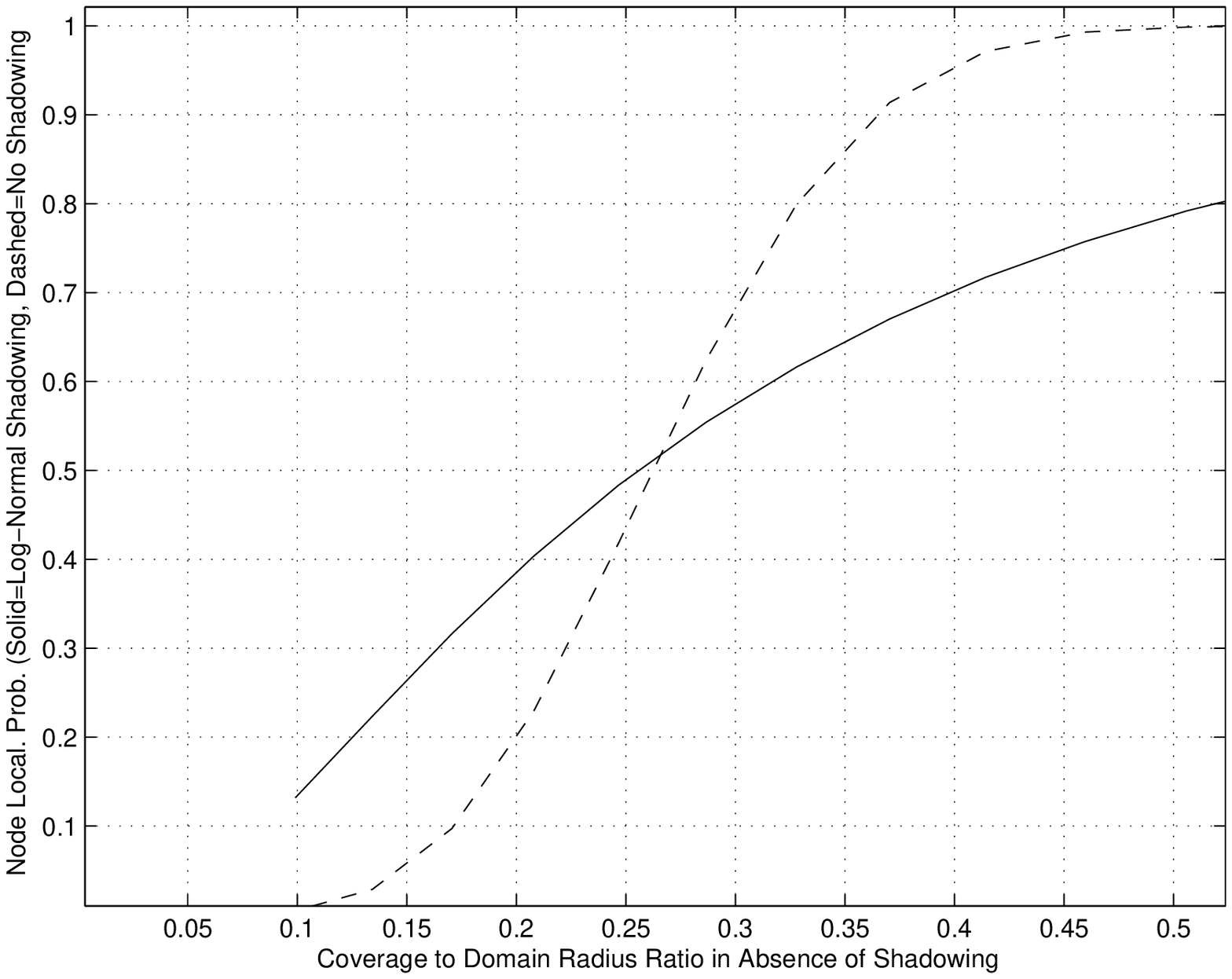}{Localization probability
as a function of $b_o$ for $a=0.2$ under both Log-Normal shadowing
(solid line) and no shadowing (dashed line).}{shadow}
Note that the limiting behavior is consistent in that if $X_s$ has
variance of zero, $\sigma_1 =0$ and the above expression reduces
to the already known result as expressed in (\ref{pf_bhat}).

For a numerical example, consider a scenario with the following
parameters: $\gamma=-80$~dBm, $P_0 =0$~dBm, $d_0 =10$~cm, $n_p
=3.5$, $\sigma_s =12$~dB, $\sigma_1 =3.43$~dB, $R=40$~m, $n=50$
and $k=10$. The computed $\hat{d}_{max}=19.3$~m and
$\hat{b}_{max}=0.48$. Fig.~\ref{shadow} depicts the localization
probability $(1-P_F)$ obtained using numerical integration for
computation of the Expectations above, as a function of $b_o$. For
comparison purposes, we also show the results obtained from
Theorem~2.1 where there is no Log-Normal shadowing. Two immediate
observations from the figure are: 1) shadowing actually improves
localization probability when $b_o$ is small, and 2) shadowing is
detrimental to localization probability when $b_o$ is relatively
large.
%
%
\subsection{Threshold Conditions}
Returning to our analysis where we assume the knowledge of the
radio coverage area of a given NL-node, a common characteristic of
many problems tackled using the probabilistic method is the
existence of a transition threshold where the characteristic of
interest exhibits a large variation. The transition threshold
observable specially for large values of $b$ can be obtained by
taking the second partial derivative of $P_F$ with respect to $a$
and setting the result to zero. Given that $P_F$ is defined
through a differential equation, it is better to work directly
with this form. This leads to the condition:
\begin{equation}
\begin{array}{ll}
\frac{(1-a)}{2}\frac{\partial^4}{\partial a^4}\left\{ [b^2 \cdot
a+(1-b^2 )]^{n-1}\right\}=&\\
\frac{\partial^3}{\partial a^3}\left\{ [b^2 \cdot a+(1-b^2
)]^{n-1}\right\}&
\end{array}
\end{equation}
The resulting closed form expression for the transition threshold
on $a$ is given by:
\begin{equation}
a^*  = 1-\frac{1}{b^2 (0.5n-1)}
\end{equation}
At the value $a^*$, there is a marked change in the localization
probability. Values of $a$ below threshold lead to a high
probability of localization of the NL-nodes. Values of $a$ greater
than $a^*$ lead to very low probability of one shot localization.
The transition threshold is plotted in Fig.~\ref{fig2} as a
function of the coverage to domain radius ratio. The noticeable
feature is the rapid increase in this threshold at a value of
$b\simeq 0.15$.

Fig.~\ref{fig3} depicts the localization probability $(1-P_F )$ as
a function of the coverage to domain radius ratio, $b$, for
exponentially increasing values of $a$ defined via the expression
$a(j)=2^{j/22}-1, \; j=1,2,...,20$ in a network containing $n=300$
nodes.

The transition threshold in this scenario can be obtained by taking the
second partial derivative of $P_F$ with respect to $b$ and setting the result to zero. This leads to
the closed form expression:
\begin{equation}
\begin{array}{ll}
b^*  = \left\{ \frac{4n^2 -n-15}{2(1-a)(2n^3 -8n^2 +10.5n -4.5)}
\left[ 1+\right.\right.&\\
\left.\left. +\sqrt{1+\frac{6(n-9)(2n^3 -8n^2 +10.5n -4.5)}{(4n^2
-n-15)^2}} \right] \right\} ^{1/2}&
\end{array}
\end{equation}
For large values of $n$, the above expression simplifies to:
\begin{equation}
b^* \simeq \left\{ \frac{1}{(1-a)n} \left[ 1+\sqrt{1.75} \right]
\right\} ^{1/2}
\end{equation}
For a fixed $a$, there is a transition threshold with respect to
$b$. At values of $b$ below $b^*$, the localization probability is
small while for values of $b$ greater than $b^*$ the localization
probability increases rapidly. The transition threshold is plotted
in Fig.~\ref{fig4} as a function of the percentage of the
NL-nodes. The noticeable feature is the relatively gradual
increase in this threshold as a function of $a$.

So far we have looked at what we may call a one-shot localization
of the NL-nodes. What we mean is that our calculated localization
probability looks at the event that a given NL-node finds at least
three L-nodes within its radio coverage. Once a NL-node localizes
itself, technically, it could become a reference localization node
and change from the category of NL-node, to the category of
L-node. The process of NL-node localization can then repeat itself
with the potentially newly added L-nodes. The study of this
iterative localization process is beyond the scope of the present
work. The asymptotic behavior however in the limiting case of
infinite iterations is relatively easy to deduce. In particular,
there is always a non-zero probability that at least one node may
not get localized even if all the other nodes are L nodes. This
provides the lower bound:
\[
P_F(\infty) \geq \sum_{k=0}^{2}\left( \begin{array}{c} n-1 \\ k
\end{array}\right) b^{2k}[1-b^2]^{n-1-k}.
\]
\section{Simulation Results}
We have simulated the node localization problem as outlined in the previous section. The details
of our simulation setup is as follows.
\figura{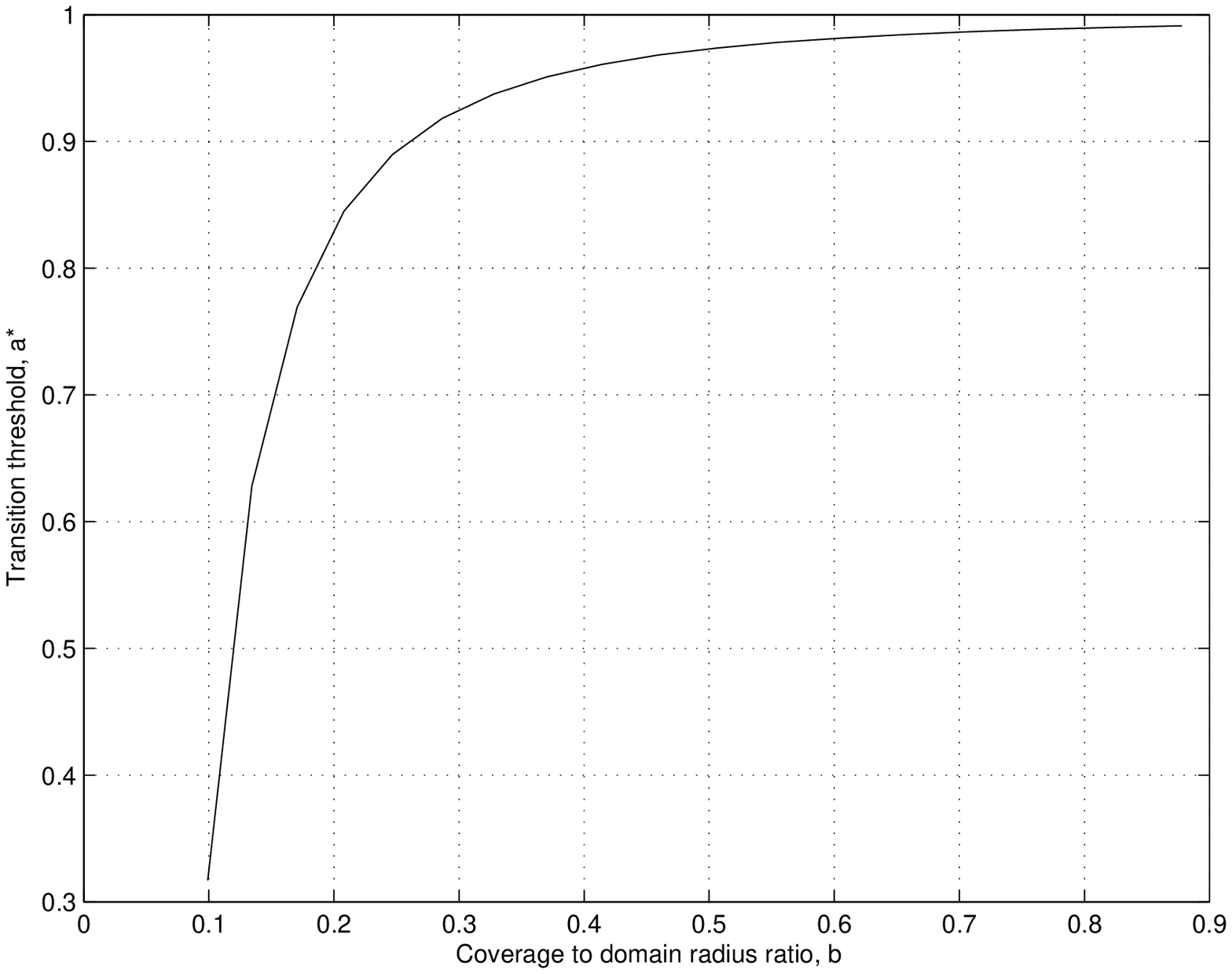}{Probability transition
threshold $a^* $ as a function of $b$ for a network containing
$n=300$ nodes.}{fig2}
\begin{itemize}
\item We consider uniformly distributing $n$ nodes over a domain
of radius $R$ normalized to $R=1$. To achieve this, we generate
$n$ pairs of uniformly distributed independent random variables
$(r_i ,\theta_i )$. Random variable $r_i$ is uninform in the
interval $[0,1]$ while $\theta_i$ is uniform in the interval
$[-\pi ,\pi ]$. To get a uniform distribution of nodes over the
domain, the variable $r_i$ is transformed to $\sqrt{r_i }$ such
that the pair $(\sqrt{r_i} ,\theta_i )$ denotes the polar
coordinates of the $i$-th node within our domain.
\item $k$ of the
$n$ nodes are identified as the L-nodes by selecting at random a
set of $k$ unique indices in the range $\{1,2,...,n\}$.
\item For
a given NL-node, the set of distances to the $k$ L-nodes are
calculated and the number of L-nodes (if any) that fall within
the coverage radius $d$ is obtained. This is repeated for all the
NL-nodes and the count of the total number of NL-nodes that can
localize themselves is obtained.
\item To get good statistical
averages, we have generated 1000 realizations of the localization
problem for a given set of parameters, $n,k,a,b$. We have then
obtained the empirical estimate of node localization failure
probability by averaging the results over 1000 realizations.
\end{itemize}
\figura{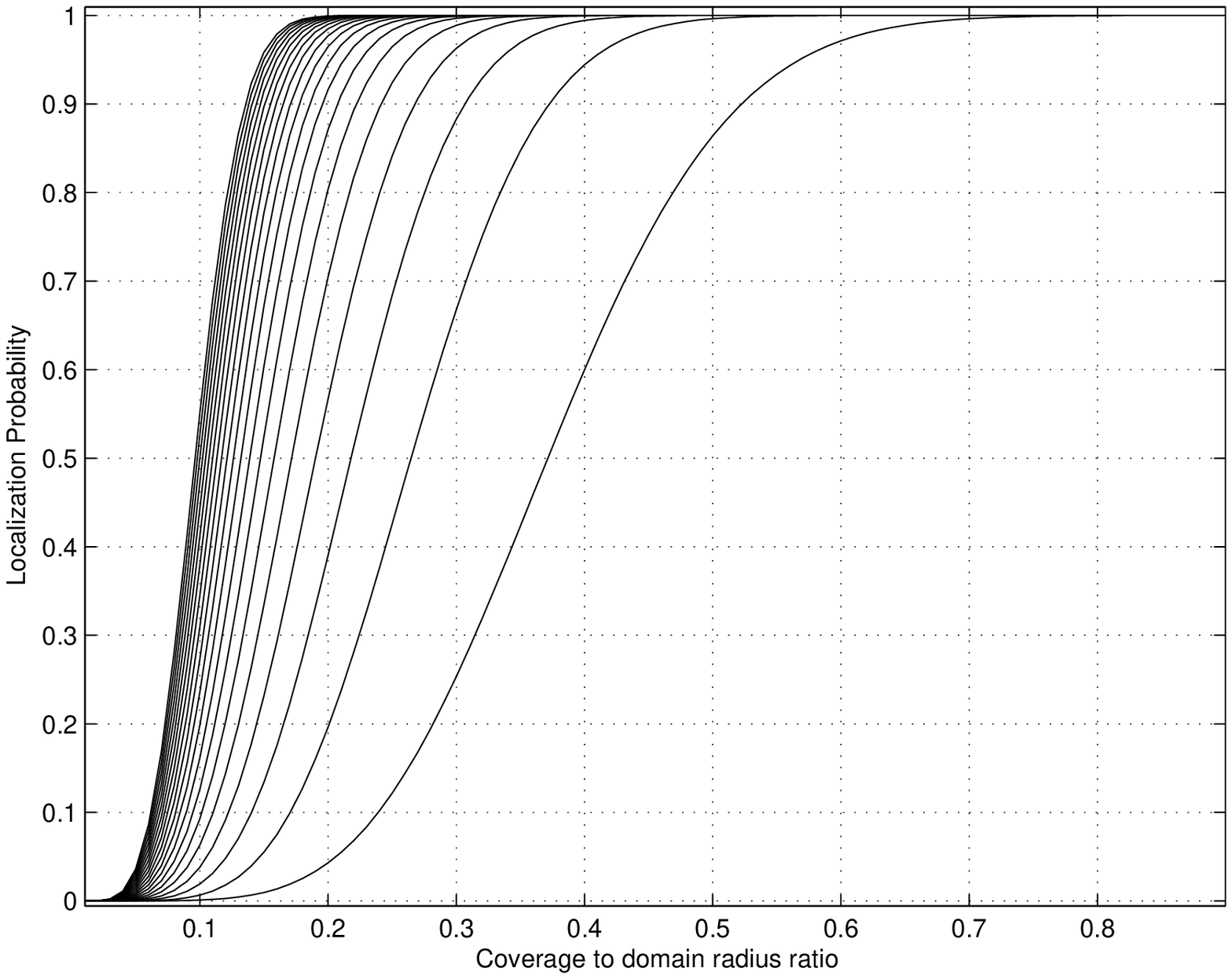}{Localization probability
as a function of $b$ for various values of $a$ in the set
$a(j)=2^{j/22}-1, \; j=1,2,...,20$ for a network containing
$n=300$ nodes. The leftmost curve is associated with the smallest
$a$. Increasing $a$ monotonically decreases the localization
probability for a given $b$.}{fig3} As an example of sample
simulation results while assessing the impact of the density of
nodes over the domain, Fig.~\ref{fig6} depicts the NL-node
localization probability versus $a$ for three different values of
$n$ at a value of $b=0.05$. Simulation points are marked by "*"
while the analytical curves obtained using the formulas derived in
the previous section are shown as solid lines. Note that at
$n=500$ the curve is convex, at $n=1000$ it is almost linear and
at $n=3000$ it is almost concave. The other noticeable effect is
the reduction in the localization probability for a fixed $a$ as
the density of the nodes within the domain is reduced. The gap
between theory and simulation can be attributed to the boundary
points which as stated previously, have an effectively smaller
coverage area than the interior points.
\figura{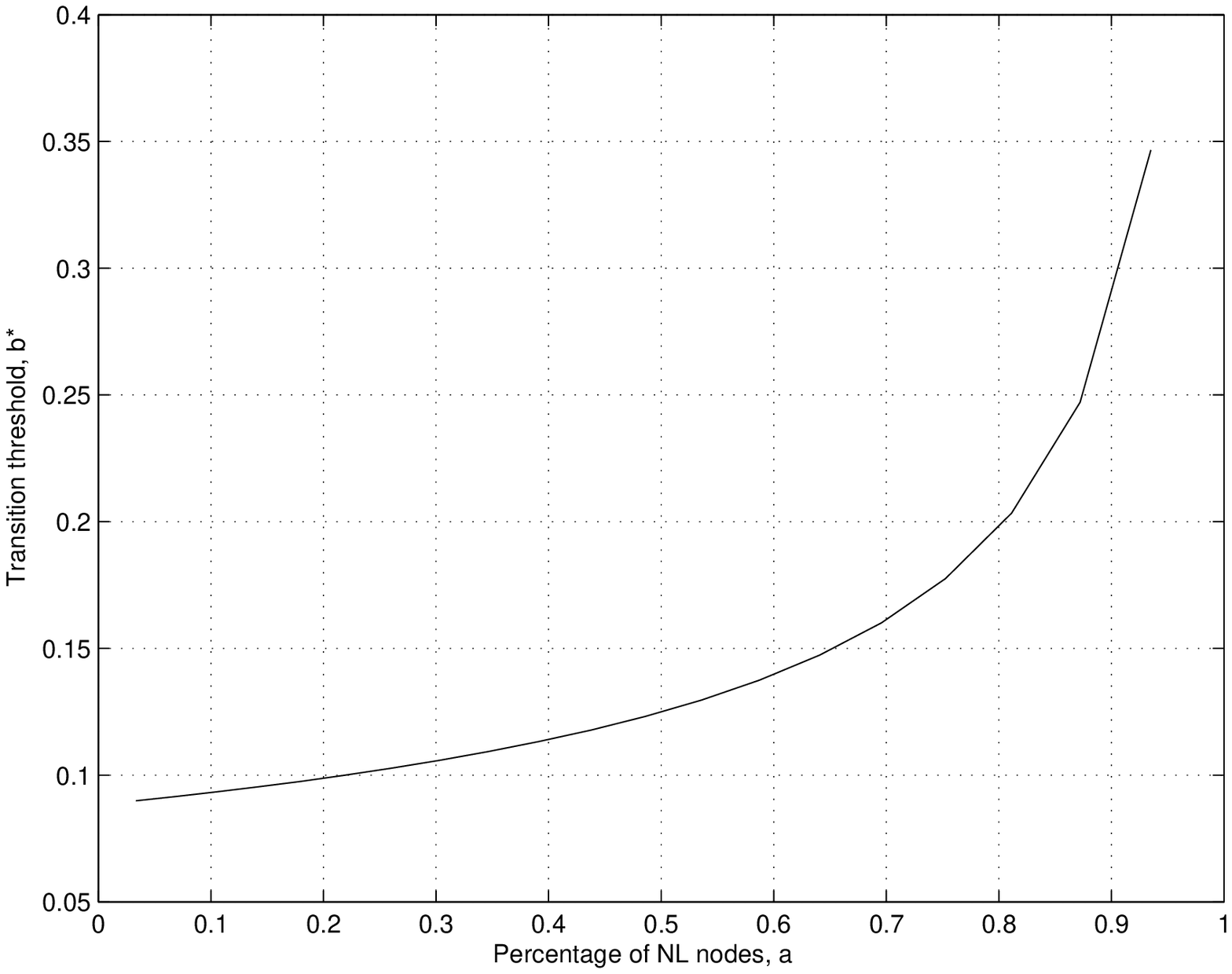}{Probability transition
threshold $b^*$ as a function of $a$ for a network containing
$n=300$ nodes.}{fig4}
\section{Conclusions}
In this article we have presented a probabilistic setup for NL-node
localization in a randomly distributed network of mixed localized
and non-localized nodes. The probabilistic method is then used to
answer some fundamental questions regarding the feasibility of
NL-node localization in such a network based on some basic
parameters such as the density of the nodes and coverage to domain
radius ratios. We have derived expressions for transition
thresholds with respect to several key parameters whereby marked
change in the localization probability is observed.
\figura{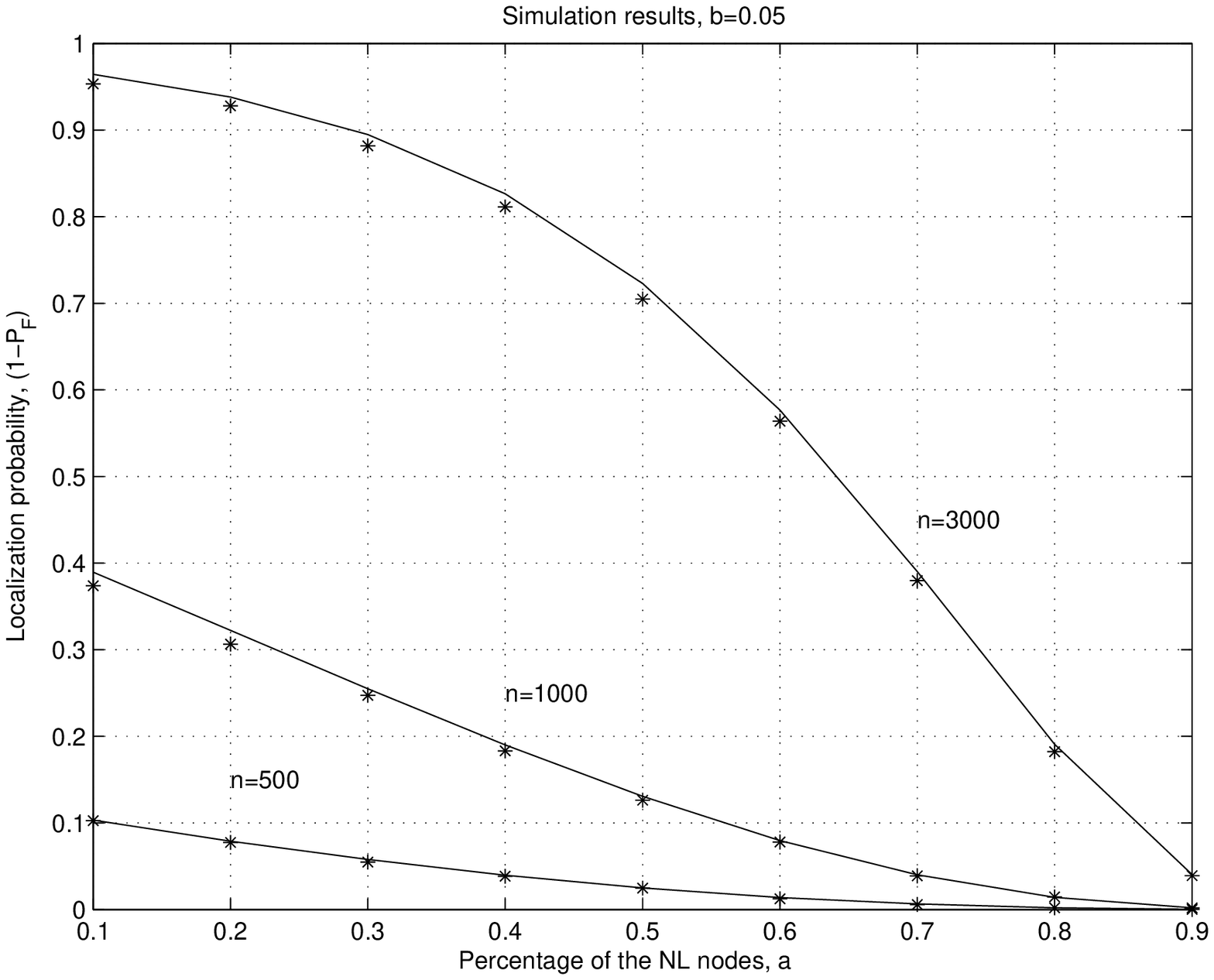}{Simulated and
theoretical $(1-P_F)$ as a function of percentage of NL-nodes $a$
(simulated points are marked by "*") for three different values of
$n$ at $b=0.05$.}{fig6}
\end{document}